\documentclass[aps, prd, twocolumn,print, showpacs, nofootinbib]{revtex4}
\usepackage{graphicx}
\begin{document}
\title{Reply to ``are GRB 090423-like bursts from the superconducting cosmic strings?"}

\maketitle

Cusps of superconducting cosmic strings were first suggested in
\cite{Ba87} as central engines driving gamma-ray bursts (GRBs). A
more elaborate description can be found in \cite{Be01}. In the
framework of such a cosmic string GRB (CSGRB) model, recently we
have shown that some high redshift GRBs (e.g., GRBs 080913 and
090423) can be well accounted for in the aspects of their
luminosities, durations, as well as their inferred star formation
rates \cite{cheng10}. In our calculations, an important angle as
$\theta\sim 1/\gamma\sim10^{-3}$ is invoked, where $\gamma$ is the
Lorentz factor of the string segment that is responsible for the GRB
prompt emission. However, in the Comment \cite{wang11} Wang, Fan, \&
Wei claim that such a very small angle could be in contradiction
with the opening angle of the GRB outflow as $\theta_{\rm
GRB}\sim0.2$, which is inferred from the GRB afterglow observations
\cite{cha10}. Although it is a very good attempt to find more
constraints on the CSGRB model from afterglow emission, it still
needs to be noticed that \textit{the angle $\theta$ acually is not
the opening angle of the GRB outflow, but is just the collimation
angle of the radiation of the corresponding string segment}. In
fact, the CSGRB model never requires that the opening angle should
be as small as $10^{-3}$. As shown in Figure 1, the GRB outflow
could instead be very wide, since all parts of the string near the
cusp can generate electromagnetic (EM) wave radiation.

In more details, since the part farther from the cusp has smaller
Lorentz factor (and thus larger radiation collimation angle), the
released EM wave energy per unit solid angle should decrease with
increasing viewing angle (the angle between the line of sight and
the direction of the string velocity at cusp) as \cite{Wi86}
\begin{equation}
{dE_{\rm em}\over d\Omega}=\left\{
\begin{array}{ll}
k{I_0^2l}/(c^2\theta^{3}),&~~{\rm for}~~\theta>\theta_{\min}\\
k{I_0^2l}/(c^2\theta_{\min}^{3}),&~~{\rm for}~~\theta\leq\theta_{\min}\\
\end{array}\right.
\end{equation}
where $\theta_{\min}\sim10^{-8}\alpha_{-8}B_{0,-7}f_z^{1/2}$ can be
determined by the equation $P_{\max}\sim{I_{\max}^2/ c}\sim{\mu l
c^2/ T_l}$. Due to the low frequencies of the EM waves, the released
energy would be absorbed by the surrounding medium and thus a
relativistic GRB outflow is produced. \textit{In view of the
string-like structure of the central engine, the structure of the
GRB outflow is likely to be an arc rather than a usually-considered
spherical cap.} The arc can be considered to be consisted by a
series of `bullets', all of which have the similar direction of
motion, as illustrated in Figure 1. The bullets can be described by
an initial Lorentz factor $\gamma_i$ and energy $E_b=\gamma_i
kI_0^2l/c^2$, where we assume that the Lorentz factor of a bullet is
same to the Lorentz factor of the corresponding portion of the
string, which is a basic assumption in the CSGRB model.
\begin{figure}
\resizebox{\hsize}{!}{\includegraphics{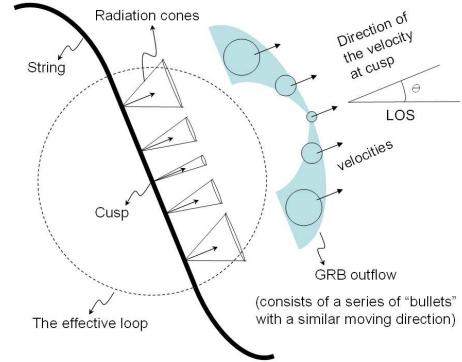}} \caption{A
schematic cartoon for the CSGRB model.}
\end{figure}

Then a more interesting question arises as whether the particular
CSGRB outflow produces the observed afterglow emission.  For a
single bullet, its sideways expansion could be insignificant, since
its adjacent bullets have the same motion direction. Denoting the
cross section of the bullet by $\mathcal S$, the dynamic evolution
of the bullet can be determined by the equation
$E_b=\gamma^2\mathcal S\mathcal Ln_1m_pc^2$, where $\mathcal
L=2\gamma^2ct$ is displacement of the bullet and $t$ is the
observer's time. Consequently, we can obtain $\gamma=\left({E_b/
2\mathcal Sn_1m_pc^3t}\right)^{1/4}\propto t^{-1/4}$. In such a
case, the afterglow emission can be roughly estimated to be
$F\propto({\epsilon_e E_b/ t})\theta_{\rm rad}^{-2}\sim{\epsilon_e
\gamma^2E_b/ t}\propto t^{-1.5}$, where $\epsilon_e$ is the electron
energy equipartition factor. On the other hand, as the deceleration
of the bullets, the later afterglow emission will be contributed by
more bullets with larger initial Lorentz factor (and thus more
energy). If these bullets are obviously separated, we may see a
series of re-brightenings in the afterglow light curves, as argued
in \cite{wang11}. However, the outflow actually is continuous. So,
instead of the re-brightenings, a smooth afterglow light curve can
be obtained, which is probably much flatter than $t^{-1.5}$. Such a
result in principle do not contradict with the afterglow observation
for GRB 090423 \cite{cha10}. Of course, a more detailed calculation
and a fitting to the observations may provide much more solid
arguments.

K. S. Cheng, Y. W. Yu, and T. Harko
\\Department of Physics, The University
of Hong Kong, Pokfulam Road, Hong Kong, China


\begin{thebibliography}{99}
\bibitem{Ba87}  A. Babul, B. Paczynski, and D. Spergel, Astrophys. J. {\bf 316},  L49
(1987); B. Paczynski, Astrophys. J. {\bf 335}, 525
(1988).
\bibitem{Be01} V. Berezinsky,  B. Hnatyk, and A. Vilenkin, Phys. Rev. {\bf D64}, 043004  (2001); V. Berezinsky,  B. Hnatyk, and A. Vilenkin, Baltic Astronomy {\bf 13}, 289
(2004)
\bibitem{cheng10} K. S. Cheng, Y. W. Yu, \& T. Harko, Phys.
Rev. Lett. {\bf 104}, 241102 (2010)
\bibitem{wang11}Y. Wang, Y. Z. Fan, \& D. M. Wei, preceding Comment, Phys. Rev. Lett.
\bibitem{cha10} P. Chandra et al. ApJ, {\bf 712}, L31 (2010)
\bibitem{Wi86} E. M. Chudnovsky, G. B. Field, D. N. Spergel, and A. Vilenkin, Phys. Rev. {\bf D34}, 944 (1986);
A. Vilenkin and T. Vachaspati, Phys. Rev. Lett. {\bf 58}, 1041
(1987)




\end{thebibliography}
\end{document}